\documentclass[draft,jgrga]{agu2001}
\usepackage{graphicx}
\authorrunninghead{LIU ET AL.}

\titlerunninghead{Global Structure of MCs}

\authoraddr{Y. Liu,
Kavli Institute for Astrophysics and Space Research, Massachusetts
Institute of Technology, Room 37-676a, Cambridge, MA 02139, USA.
Also at State Key Laboratory of Space Weather, Center for Space
Science and Applied Research, Chinese Academy of Sciences, P.O. Box
8701, Beijing 100080, China. (liuxying@space.mit.edu)}

\begin{document}

%
%
%
\setkeys{Gin}{draft=false}
%

%
%
\title{Constraints on the Global Structure of Magnetic Clouds: Transverse Size
and Curvature}


\authors{Y. Liu\altaffilmark{1, 2},
 J. D. Richardson\altaffilmark{1, 2},
 J. W. Belcher\altaffilmark{1},
 C. Wang\altaffilmark{2},
 Q. Hu\altaffilmark{3},
 and J. C. Kasper\altaffilmark{1}}

 \altaffiltext{1}{Kavli Institute for Astrophysics and Space Research,
 Massachusetts Institute of Technology, Cambridge, MA, USA}

 \altaffiltext{2}{State Key Laboratory of Space Weather, Center
for Space Science and Applied Research, Chinese Academy of Sciences,
P.O. Box 8701, Beijing 100080, China}

 \altaffiltext{3}{Institute for Geophysics and Planetary Physics,
 University of California, Riverside, CA, USA}

%
%

\begin{abstract}
We present direct evidence that magnetic clouds (MCs) have highly
flattened and curved cross section resulting from their interaction
with the ambient solar wind. Lower limits on the transverse size are
obtained for three MCs observed by ACE and Ulysses from the
latitudinal separation between the two spacecraft, ranging from
40$^{\circ}$ to 70$^{\circ}$. The cross-section aspect ratio of the
MCs is estimated to be no smaller than $6:1$. We offer a simple
model to extract the radius of curvature of the cross section, based
on the elevation angle of the MC normal distributed over latitude.
Application of the model to Wind observations from 1995 - 1997
(close to solar minimum) shows that the cross section is bent
concavely outward by a structured solar wind with a radius of
curvature of $\sim$ 0.3 AU. Near solar maximum, MCs tend to be
convex outward in the solar wind with a uniform speed; the radius of
curvature is proportional to the heliographic distance of MCs, as
demonstrated by Ulysses observations between 1999 and 2003. These
results improve our knowledge of the global morphology of MCs in the
pre-Stereo era, which is crucial for space weather prediction and
heliosphere studies.
\end{abstract}

%
%

%

\begin{article}

\section{Introduction}
Coronal mass ejections (CMEs) are spectacular eruptions in the solar
corona. In addition to 10$^{15-16}$ g of plasma, CMEs carry a huge
amount of magnetic flux and helicity into the heliosphere. Their
interplanetary manifestations (ICMEs) often show a regular magnetic
pattern; ICMEs with this pattern have been classified as magnetic
clouds (MCs). MCs are characterized by a strong magnetic field, a
smooth and coherent rotation of the magnetic field vector, and a
depressed proton temperature compared to the ambient solar wind
[Burlaga et al., 1981].

MCs drive many space weather events and affect the solar wind
throughout the heliosphere, so it is important to understand their
spatial structure. Most in situ observations give information on a
single line through an MC; flux-rope fitting techniques have been
developed to interpret these local measurements. Cylindrically
symmetric models vary from a linear force-free field [e.g., Burlaga,
1988; Lepping et al., 1990] to non-force-free fields with a current
density dependence [e.g., Hidalgo et al., 2002a; Cid et al., 2002].
Elliptical models take into account the expansion and distortion
effect of MCs, also based on the linear force-free [Vandas and
Romashets, 2003] and non-force-free approaches [e.g., Mulligan and
Russell, 2001; Hidalgo et al., 2002b]. The Grad-Shafranov (GS)
technique relaxes the force-free assumption and reconstructs the
cross section of MCs in the plane perpendicular to the cloud's axis
without prescribing the geometry [e.g., Hau and Sonnerup, 1999; Hu
and Sonnerup, 2002]. Although useful in describing local
observations, these models may significantly underestimate the true
dimension, magnetic flux and helicity of MCs [Riley et al., 2004;
Dasso et al., 2005]; the ambiguities in their results cannot be
removed since they involve many free parameters and assumptions.
Multiple point observations are therefore required to properly
invert the global structure of MCs.

Indirect evidence, both from observations and numerical simulations,
suggests that MCs are highly flattened and distorted due to their
interaction with the ambient solar wind. CMEs observed at the solar
limb typically have an angular width of 50 - 60$^{\circ}$ and
maintain this width as they propagate through the corona [e.g., Webb
et al., 1997; St. Cyr et al., 2000]. At 1 AU, this angular width
would correspond to a size of $\sim$ 1 AU, much larger than the
ICME's radial thickness of $\sim$ 0.2 AU [e.g., Liu et al., 2005;
Liu et al., 2006a]. Shocks driven by fast MCs have a standoff
distance which is too large to be produced by a cylindrically
symmetric flux rope [Russell and Mulligan, 2002]. The oblate cross
section of MCs is also indicated by global magnetohydrodymic (MHD)
simulations of the propagation both in a uniform [e.g., Cargill et
al., 2000; Odstrcil et al., 2002; Riley et al., 2003] and structured
solar wind [e.g., Groth et al., 2000; Odstrcil et al., 2004;
Manchester et al., 2004].

The simulated flux ropes show an interesting curvature which depends
on the background solar wind state. Figure~1 shows an idealized
sketch of flux ropes in the solar meridianal plane, initially having
a radius $r_0=1$ $r_{\odot}$ at a height $h_0=2$ $r_{\odot}$ from
the Sun, where $r_{\odot}$ represents the solar radius. This
configuration corresponds to an angular extent of
$\Delta\theta=60^{\circ}$ subtended by the rope. At a time $t$, the
axis-centered distance $r$ and polar angle $\phi$ in the flux-rope
cross section translate to the heliographic distance ($R$) and
latitude ($\theta$) assuming kinematic evolution [Riley and Crooker,
2004; Owens et al., 2006]
$$\theta = \arctan({r\sin\phi \over h_0 + r\cos\phi}),$$ and
$$ R = \sqrt{r^2\sin^2\phi + (h_0+r\cos\phi)^2} + vt[1+{Ar \over
r_0}\cos(\phi-\theta)],$$ where $v$ is the solar wind speed and $A =
0.1$ is the ratio of the expansion speed at the rope edge (relative
to the rope center) to the solar wind speed. The left flux rope is
propagating into a uniform solar wind with a speed of $v=450$ km
s$^{-1}$, while the right one is propagating into a solar wind with
a latitudinal speed gradient $v=700 \sin^2\theta + 400$ km s$^{-1}$.
After 4 days these flux ropes arrive at 1 AU. Due to the expansion
of the solar wind, plasmas on different stream lines move apart
while the magnetic tension in the flux rope tries to keep them
together. Since the flow momentum overwhelms the magnetic force
after a few solar radii, the flux rope is stretched azimuthally but
its angular extent is conserved, as shown in Figure~1. The left case
is representative of solar maximum, when the solar wind speed is
roughly uniform in the meridianal plane. The right panel represents
solar minimum, when fast solar wind originates from large polar
coronal holes and slow wind is confined to low latitudes associated
with helmet streamers [e.g., McComas et al., 1998]. The flux rope is
bent convexly outward by a spherical expansion of the solar wind
(left panel) or concavely outward by a structured wind (right
panel).

\begin{figure*}
\centerline{\includegraphics[width=35pc]{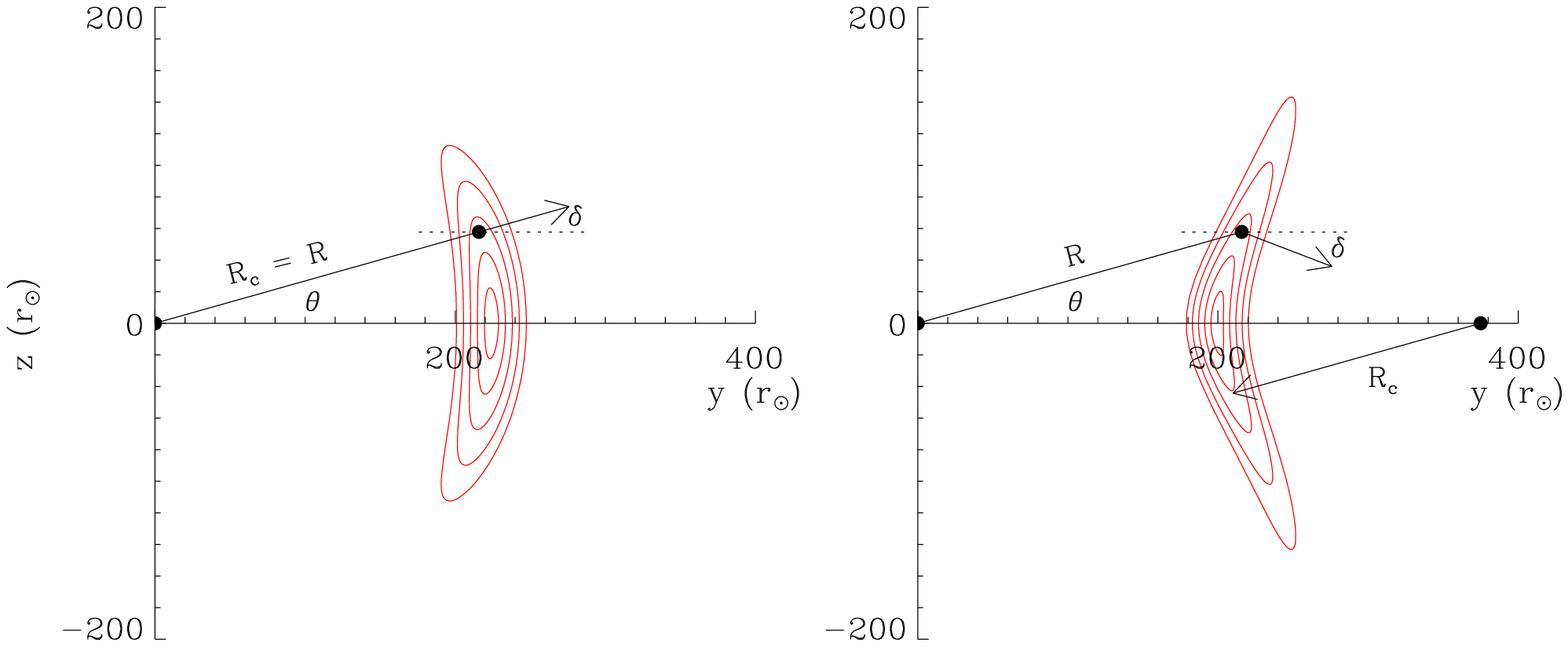}}
\caption{Schematic diagram of MCs at 1 AU in the solar meridianal
plane with axes perpendicular to the radial and transverse
directions, illustrating the large latitudinal extent and curvature
in a uniform (left panel, corresponding to solar maximum) and
structured solar wind (right panel, corresponding to solar minimum).
Contours denote levels of the initial flux-rope radius. The angles,
labeled as $\theta$ and $\delta$, represent the latitude of a
virtual spacecraft and the elevation angle of the flux-rope normal.
The distance of the spacecraft and radius of flux-rope curvature are
marked as $R$ and $R_c$, respectively.}
\end{figure*}

From Figure~1, we obtain a simple relationship between the latitude
$\theta$ of an observing spacecraft and the normal elevation angle
$\delta$ of the flux rope at the spacecraft
\begin{equation}
\delta = \arcsin({R \over R_c}\sin\theta).
\end{equation}
The radius of curvature, $R_c$, is defined such that it is positive
when the flux rope is curved away from the Sun (left panel) and
negative when curved toward the Sun (right panel). In the left case,
$R_c = R$, so equation~1 is reduced to $\theta=\delta$. Since MCs
are highly flattened as discussed above, the normal would be along
the minimum variance direction of the flux-rope magnetic field. The
radius of curvature of MCs can be extracted from equation~1 by
examining the latitudinal distribution of the normal elevation
angles. Lower limits for the transverse size of MCs can be derived
from pairs of spacecraft widely separated in latitude. Ulysses,
complemented with a near-Earth spacecraft, is particularly useful
for this research since it covers latitudes up to 80$^{\circ}$
[e.g., Hammond et al., 1995; Gosling et al., 1995].

This paper applies the above methodology to give the first direct
observational evidence for the large-scale transverse size and
curvature of MCs. The data and analysis methods are described in
section~2. Sections~3 and 4 give lower limits for the transverse
size of MCs and study how they are curved in different solar wind
states, respectively. We summarize and discuss the results in
section~5.

\section{Observations and Data Analysis}

\begin{figure*}
\includegraphics[width=18pc]{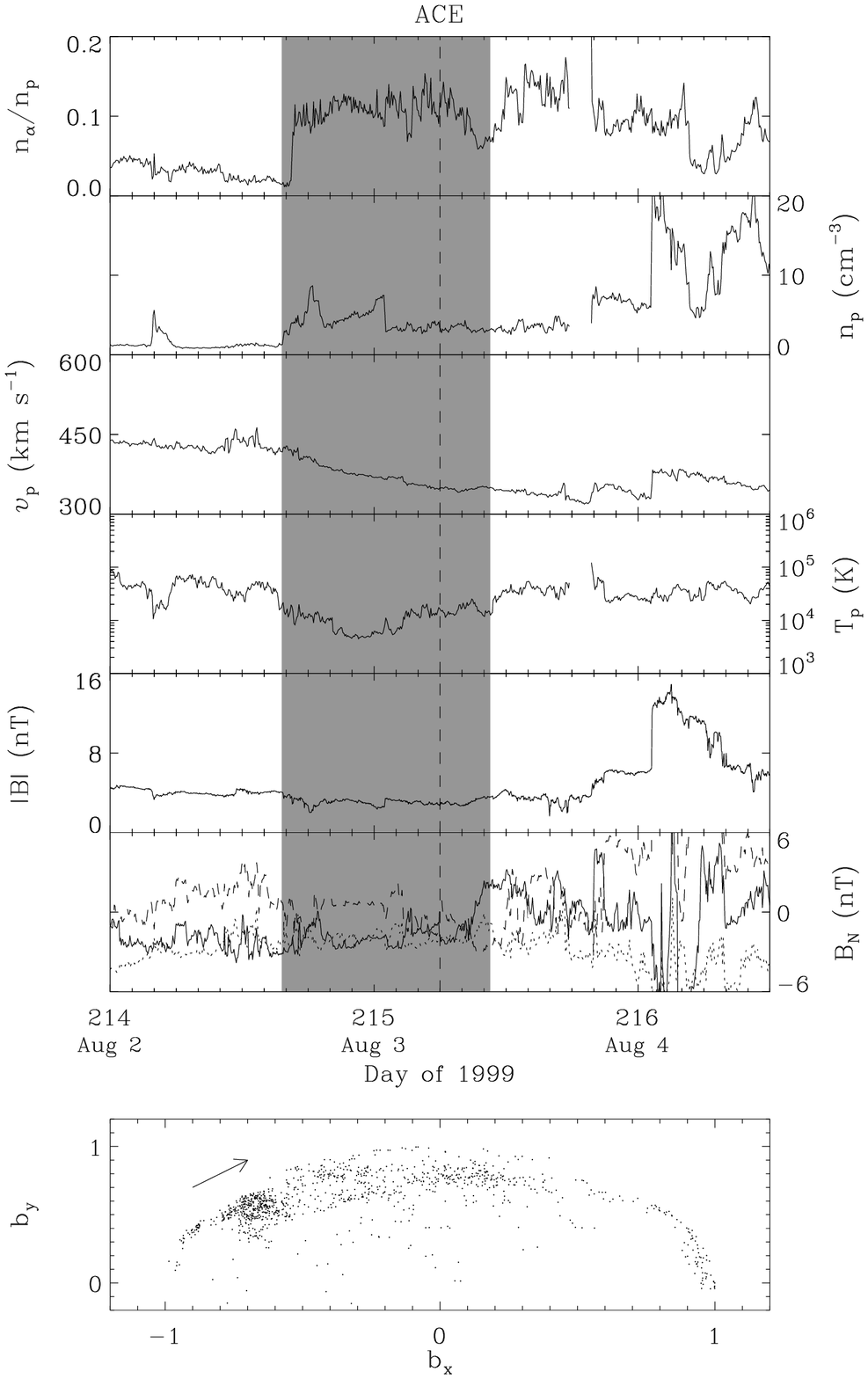}
\includegraphics[width=18pc]{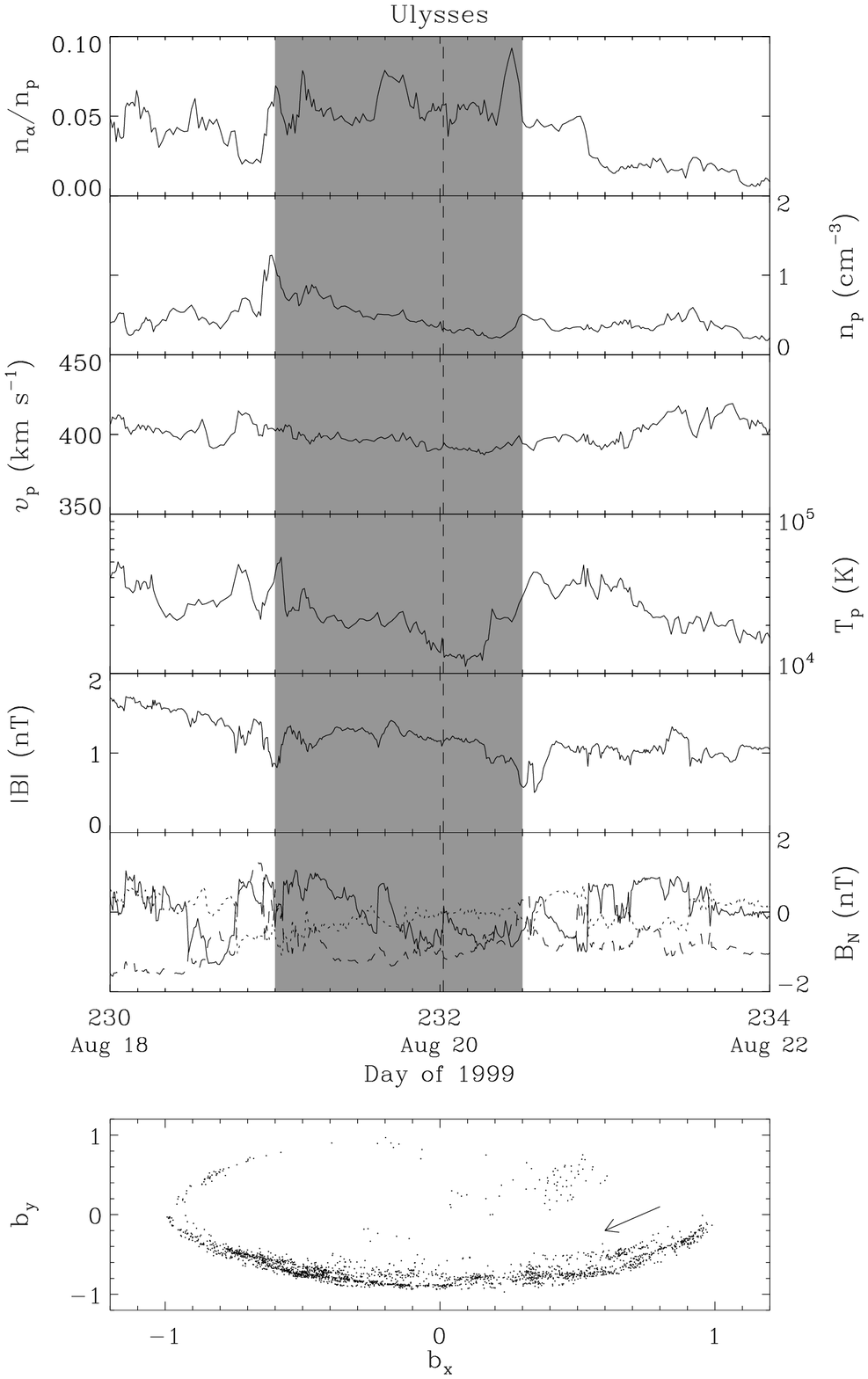} \caption{Solar wind plasma
and magnetic field parameters measured by ACE (left panel) and
Ulysses (right panel) for Case 1 in Table~1. From top to bottom, the
panels show the alpha-to-proton density ratio, proton density, bulk
speed, proton temperature, magnetic field strength, magnetic
components (solid line for $B_N$, dotted line for $B_R$, dashed line
for $B_T$), and rotation of the normalized magnetic field vector
inside the MC in the maximum variance plane. The shaded region shows
the MC. Dashed lines denote the separation of the two flux ropes
contained in the MC. Arrows in the bottom panels show the direction
of the magnetic field rotation.}
\end{figure*}

\begin{figure*}
\includegraphics[width=18pc]{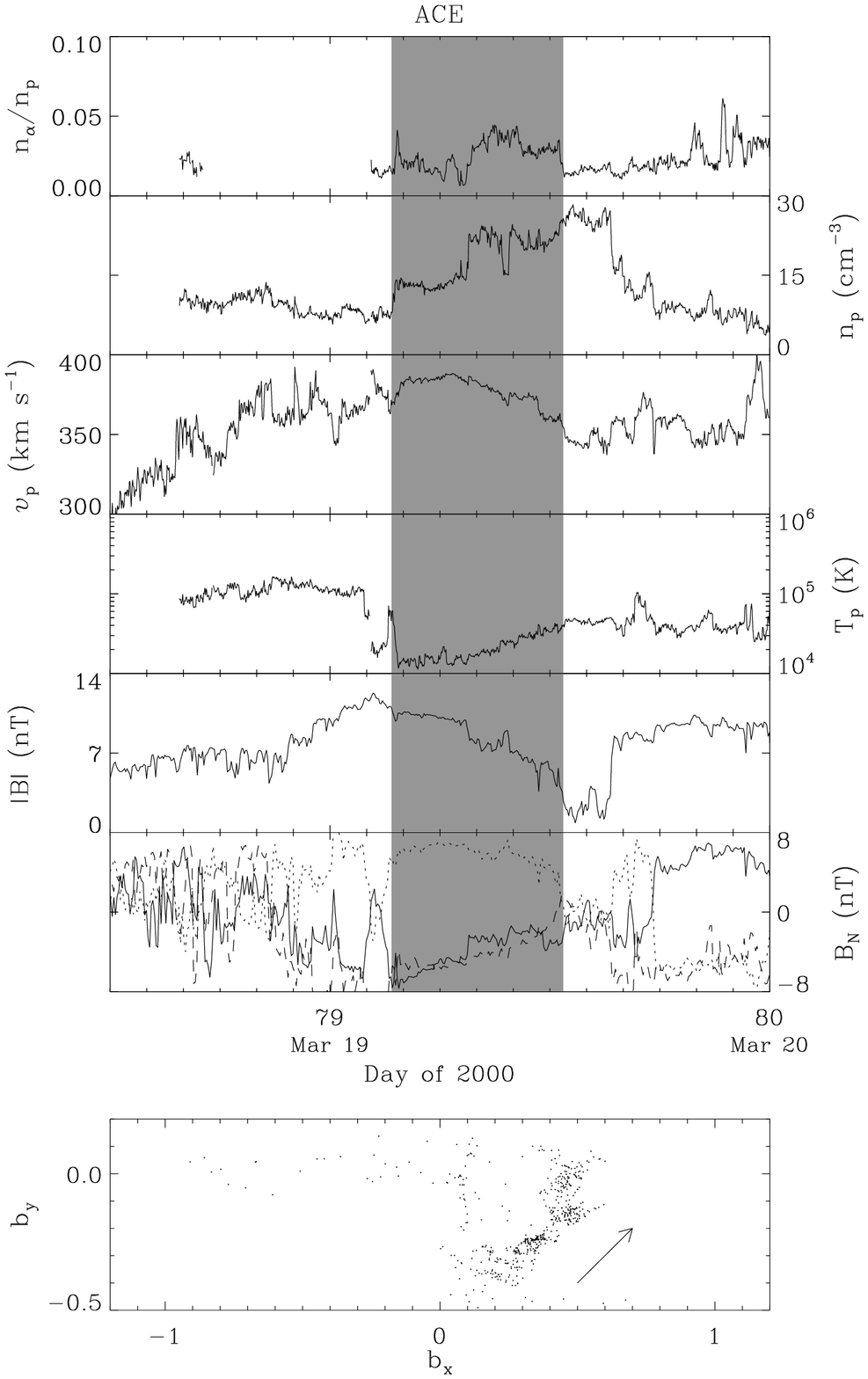}
\includegraphics[width=18pc]{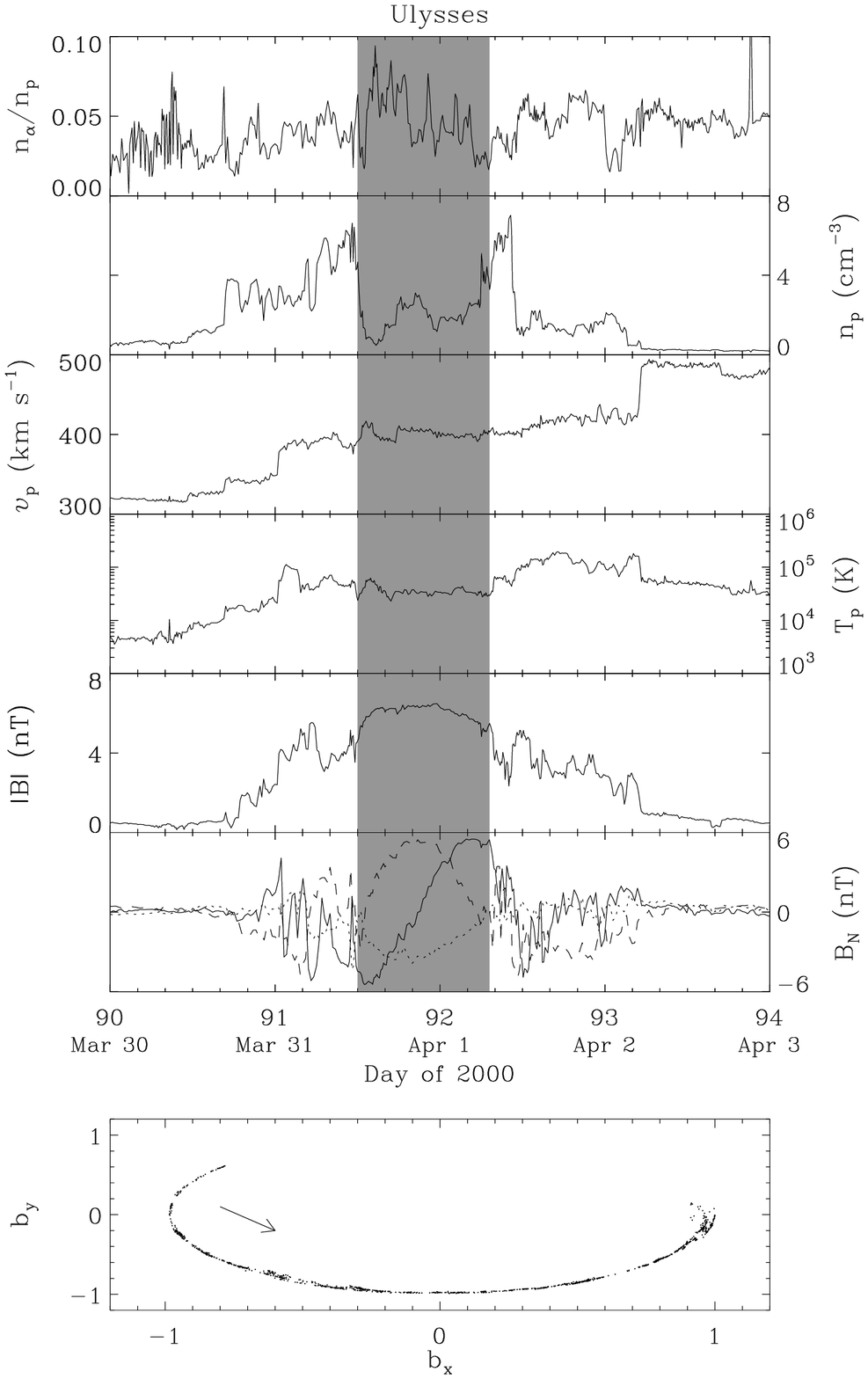} \caption{Solar wind plasma
and magnetic field parameters for Case 2 in Table~1. Same format as
Figure~2.}
\end{figure*}

\begin{figure*}
\includegraphics[width=18pc]{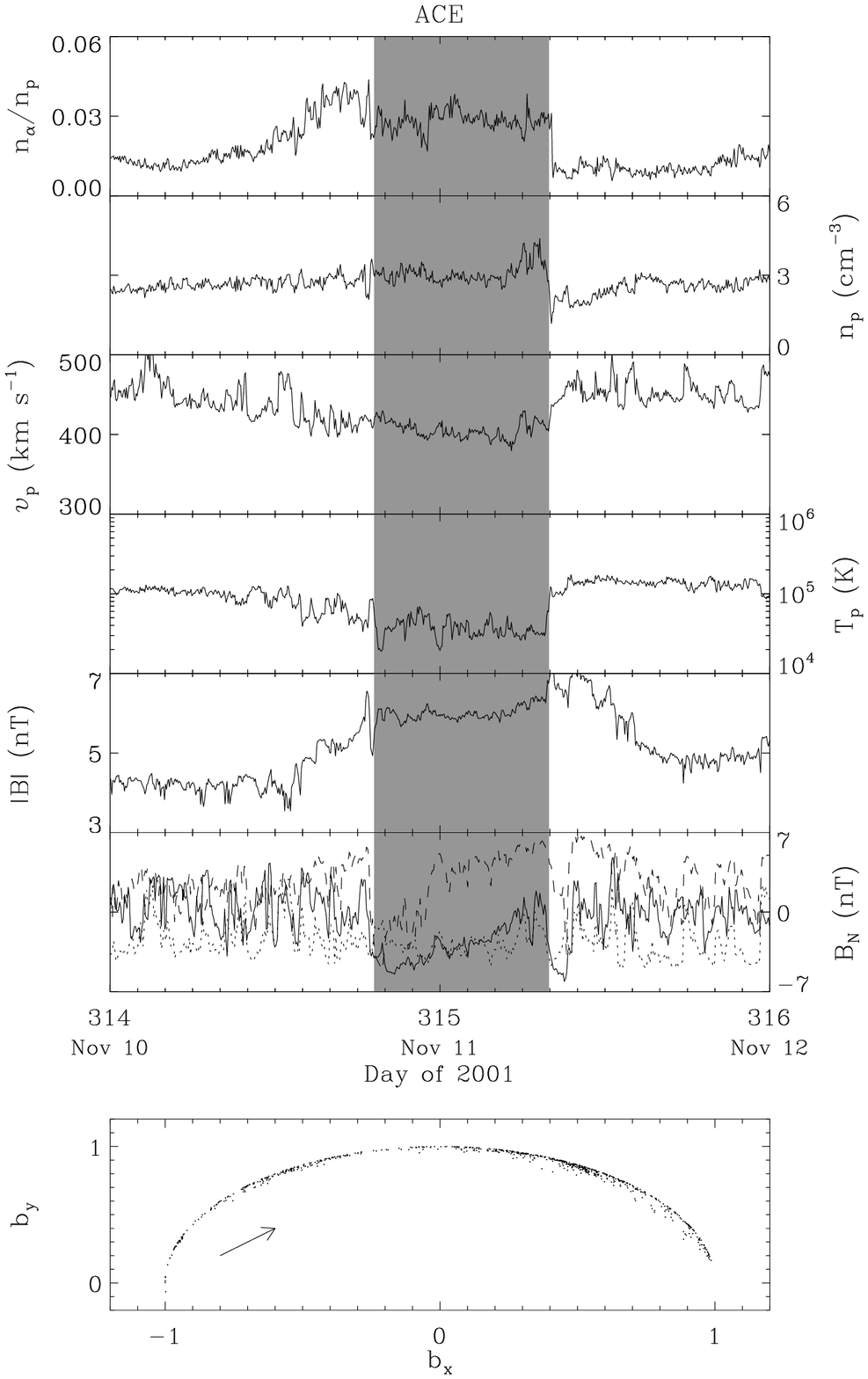}
\includegraphics[width=18pc]{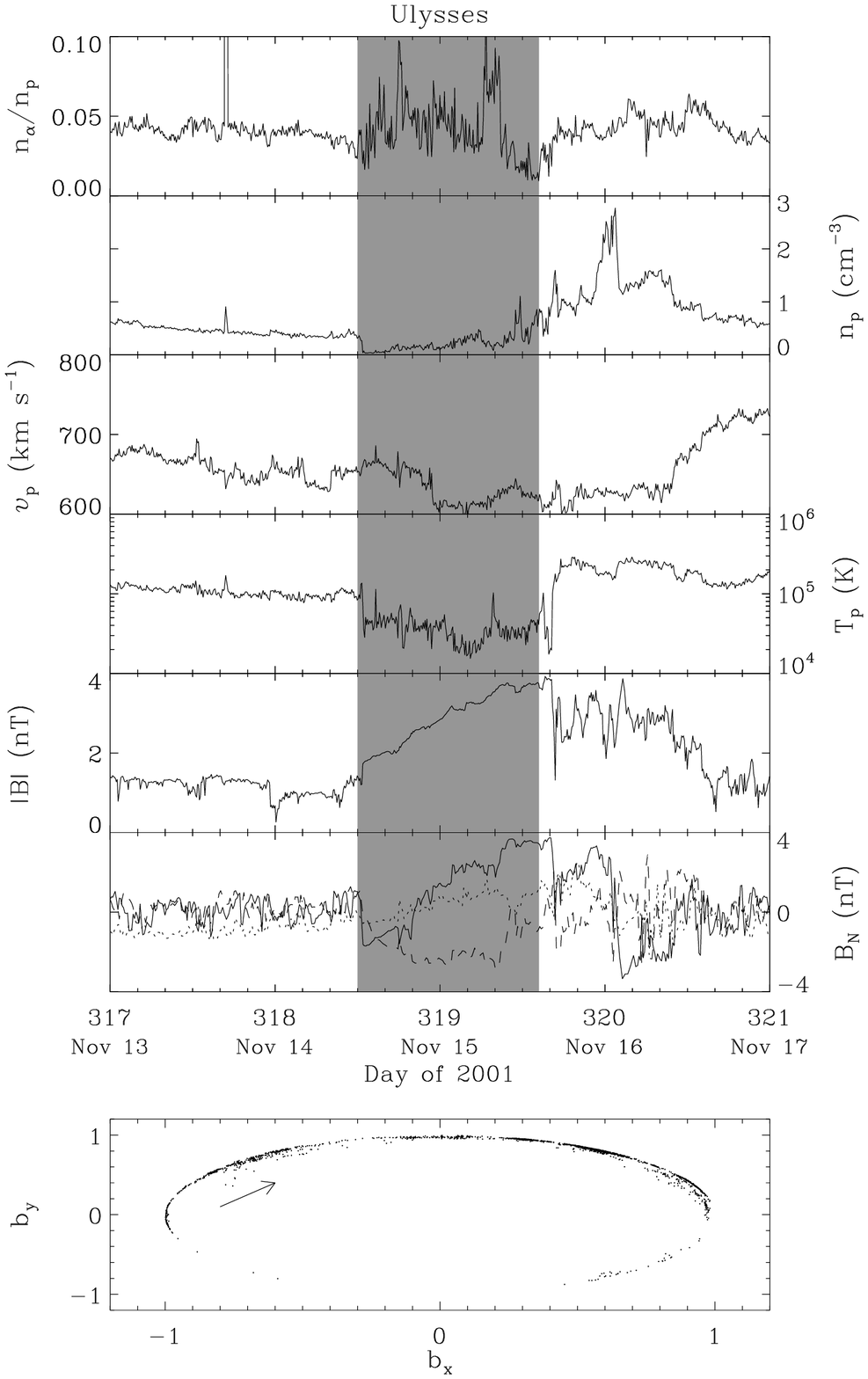} \caption{Solar wind plasma
and magnetic field parameters for Case 3 in Table~1. Same format as
Figure~2.}
\end{figure*}

To give a meaningful measure of the transverse size, we need at
least two spacecraft separated as widely as possible in the solar
meridianal plane. Launched in 1991, Ulysses explores the solar wind
conditions at distances from 1 to 5.4 AU and up to 80$^{\circ}$ in
latitude. Wind and ACE have provided near-Earth measurements (within
7$^{\circ}$ of the solar equatorial plane) since 1994 and 1998,
respectively. We first look for MCs in Ulysses data when it is more
than 30$^{\circ}$ away from the solar equator. If we see the same MC
at the near-Earth spacecraft, then its transverse width is at least
the spacecraft separation. To determine if the spacecraft observe
the same MC, we look at the timing and data similarities (similar
transient signatures, the same chirality, etc.), and use a
one-dimensional (1-D) MHD model to do data alignment.

We require that the MC axis lie close to the solar equatorial plane.
Otherwise, the width we invert is the length of the flux rope other
than the transverse size of the cross section; this requirement also
minimizes the effect of axial curvature. In addition, our curvature
study needs MCs not to be aligned with the radial direction. Wind
observations in 1995 - 1997 are used to investigate the MC curvature
near solar minimum, while Ulysses as well as the near-Earth
spacecraft provides observations from 1999 to 2003 for the curvature
study near solar maximum.

\subsection{Coordination of Observations via 1-D MHD Modeling}
Single-point observations can only sample MCs at a specific
distance. As MCs propagate in the solar wind, they may change
appreciably. Models are needed to connect observations at different
spacecraft. Studies to compare ICME observations at various
locations have been performed, using ICME signatures and an MHD
simulation to trace their evolution [e.g., Wang et al., 2001;
Richardson et al., 2002; Riley et al., 2003]. We follow the same
approach and use a 1-D MHD model developed by Wang et al. [2000] to
propagate the solar wind from 1 AU to Ulysses. Momentum and energy
source terms resulting from the interaction between solar wind ions
and interstellar neutrals can be included in this model, but we drop
them since they have a negligible effect on the solar wind
propagation within Ulysses' distance. All physical quantities at the
inner boundary (1 AU) are set to the average near-Earth solar wind
conditions; the MHD equations are then solved by a piecewise
parabolic scheme to give a steady state solar wind solution.
Observations at 1 AU (typically 60 days long) containing the ICME
data are introduced into the model as perturbations. The numerical
calculation stops when the perturbation has traveled to Ulysses. The
model output is compared with Ulysses observations. Note that the
1-D model assumes spherical symmetry, so we do not expect the model
output to exactly match the Ulysses data. Nevertheless, large stream
structures should be similar and allow us to align these data sets.

\subsection{Minimum Variance Analysis}
The axis orientation of MCs is needed to specify their global
structure. Minimum variance analysis (MVA) of the measured magnetic
field yields useful principle axes [e.g., Sonnerup and Cahill, 1967;
Sonnerup and Scheible, 1998]. A normal direction, $\hat{\bf n}$, can
be identified by minimizing the deviation of the field component
${\bf B}^m \cdot \hat{\bf n}$ from $\langle{\bf B}\rangle \cdot
\hat{\bf n}$ for a series of measurements of $m=1, \ldots, N$
\begin{equation}
\sigma^2 = {1 \over N}\sum_{m=1}^N |({\bf B}^m-\langle{\bf
B}\rangle) \cdot \hat{\bf n}|^2,
\end{equation}
where $\langle{\bf B}\rangle$ is the average magnetic field vector.
Optimizing the above equation under the constraint of $|\hat{\bf
n}|^2=1$ results in the eigenvalue problem of the covariance matrix
of the magnetic field
\begin{equation}
M_{\mu\nu} = \langle B_{\mu}B_{\nu}\rangle - \langle
B_{\mu}\rangle\langle B_{\mu}\rangle,
\end{equation}
where $\mu, \nu$ indicate the field components in Cartesian
coordinates. Eigenvectors of the matrix, $\hat{\bf x}^{\star}$,
$\hat{\bf y}^{\star}$, $\hat{\bf z}^{\star}$, corresponding to the
eigenvalues in order of decreasing magnitude, denote the the
maximum, intermediate and minimum variance directions. The normal of
an elongated flux rope should be along the minimum variance
direction (see Figure~1); the maximum variance would occur
azimuthally since the azimuthal component changes its sign across
the flux rope; the intermediate variance direction is identified as
the axis orientation due to the non-uniform distribution of the
axial field over the flux-rope cross section. The MVA method also
gives the chirality of the flux-rope fields as shown in the bottom
panels of Figures~2, 3 and 4.

Angular error estimates of the directions can be written as
[Sonnerup and Scheible, 1998]
\begin{equation}
\Delta\varphi_{ij} = \sqrt{{\lambda_z(\lambda_i+\lambda_j-\lambda_z)
\over (N-1)(\lambda_i-\lambda_j)^2}}
\end{equation}
for $i, j \in\{\hat{\bf x}^{\star}, \hat{\bf y}^{\star}, \hat{\bf
z}^{\star}\}$ and $i\neq j$, where $\lambda_i$ denotes the
eigenvalue of the variance matrix, and $\Delta\varphi_{ij}$
represents the angular uncertainty of eigenvector $i$ with respect
to eigenvector $j$. The uncertainty of the normal elevation angle
$\delta$ is
\begin{equation}
\Delta\varphi_z = \sqrt{(\Delta\varphi_{zx})^2
+(\Delta\varphi_{zy})^2},
\end{equation}
where we assume that the errors are independent.

\subsection{Grad-Shafranov Technique}

\begin{table*}
\caption{Estimated parameters of MCs at ACE\tablenotemark{a}~~and
Ulysses\tablenotemark{a}}
\begin{footnotesize}
\begin{tabular*}{\textwidth}{@{\extracolsep{\fill}}lcccccccccc}
\hline No.&\multicolumn{1}{c}{\vrule height 10pt width 0pt Year}
&\multicolumn{1}{c}{CME Onset\tablenotemark{b}}
&\multicolumn{1}{c}{Start} &\multicolumn{1}{c}{End}
&\multicolumn{1}{c}{$R$\tablenotemark{c}}
&\multicolumn{1}{c}{$\theta$\tablenotemark{c}}
&\multicolumn{1}{c}{$\phi$\tablenotemark{c}}
&\multicolumn{1}{c}{$\Theta$\tablenotemark{d}}
&\multicolumn{1}{c}{$\Phi$\tablenotemark{d}}
&\multicolumn{1}{c}{Chirality} \cr
~ &~ &~ &~ &~ &(AU) &($^{\circ}$) &($^{\circ}$) &($^{\circ}$)
&($^{\circ}$) &~ \cr \hline
1 &\multicolumn{1}{c}{\vrule height 10pt width 0pt 1999} &Jul 29,
05:43 &Aug 2, 15:36 &Aug 3, 10:34 &1 &5.9 &234.3 &$-$42.4 ($-$48.8)
&251.2 (261.5) &L\cr
~ &\multicolumn{1}{c}{\vrule height 10pt width 0pt 1999} &- &Aug 19,
00:00 &Aug 20, 12:00 &4.7 &$-$32.3 &87.4 &$-$7.1 ($-$11.3) &283.9
(261.9) &L\cr
2 &\multicolumn{1}{c}{\vrule height 10pt width 0pt 2000} &Mar 14,
10:08 &Mar 19, 03:22 &Mar 19, 12:43 &1 &$-$7.1 &103.0 &24.3 (29.2)
&51.4 (95.8) &R\cr
~ &\multicolumn{1}{c}{\vrule height 10pt width 0pt 2000} &- &Mar 31,
12:00 &Apr 1, 07:12 &3.7 &$-$50.1 &93.1 &19.1 (13.6) &112.2 (90.0)
&R\cr
3 &\multicolumn{1}{c}{\vrule height 10pt width 0pt 2001} &Nov 6,
16:39 &Nov 10, 19:12 &Nov 11, 07:55 &1 &3.3 &333.0 &$-$37.8
($-$16.5) &151.1 (121.9) &L\cr
~ &\multicolumn{1}{c}{\vrule height 10pt width 0pt 2001} &- &Nov 14,
12:00 &Nov 15, 14:24 &2.3 &75.4 &39.5 &$-$16.6 ($-$15.4) &104.7
(103.4) &L\cr
PR\tablenotemark{e} &\multicolumn{1}{c}{\vrule height 10pt width 0pt
1999} &Feb 15, -\tablenotemark{f} &Feb 18, 13:55 &Feb 19, 11:02 &1
&$-$7.0 &74.0 &0.4 ($-$30.1) &277.1 (284.7) &L\cr
~ &\multicolumn{1}{c}{\vrule height 10pt width 0pt 1999} &- &Mar 3,
21:36 &Mar 5, 21:36 &5.1 &$-$22.3 &85.2 &$-$38.0 ($-$35.3) &269.3
(265.2) &L\cr \hline
\end{tabular*}
\end{footnotesize}
\tablenotetext{a}{Corresponding to the first and second lines for
each case, respectively.}
\tablenotetext{b}{Obtained by extrapolating the quadratic fit of the
CME's height-time curve to the solar surface.}
\tablenotetext{c}{Heliographic inertial distance, latitude and
longitude of the spacecraft.} \tablenotetext{d}{Axis elevation angle
with respect to the solar equatorial plane and azimuthal angle in
RTN coordinates, estimated from MVA (outside parentheses) and the GS
technique (inside parentheses).}
\tablenotetext{e}{The case studied by Riley et al. [2003].}
\tablenotetext{f}{No CMEs observed at the Sun on 1999 February 15.}
\end{table*}

Initially designed for the study of the terrestrial magnetopause
[e.g., Hau and Sonnerup], the GS technique can be applied to
flux-rope reconstruction [e.g., Hu and Sonnerup, 2002]. It assumes
an approximate deHoffmann-Teller (HT) frame in which the electric
field vanishes everywhere. Structures in such a frame obey MHD
equilibrium, ${\bf j}\times {\bf B} - \nabla p = 0$, which can be
reduced to the so-called GS equation [e.g., Sturrock, 1994]
\begin{equation}
{\partial^2A \over \partial x^2} + {\partial^2A \over \partial y^2}
= -\mu_0{d \over dA}(p + {B_z^2 \over 2\mu_0})
\end{equation}
by assuming a translational symmetry along the flux rope (i.e.,
$\partial\over \partial z$=0). The vector potential is defined as
${\bf A} = A(x, y) \hat{\bf z}$, through which the magnetic field is
given by ${\bf B} = ({\partial A \over \partial y}, -{\partial A
\over \partial x}, B_z)$.

The key idea in reconstructing the flux rope is that the thermal
pressure $p$ and the axial field $B_z$ are functions of $A$ alone.
The flux-rope orientation is determined by the single-valued
behavior of the transverse pressure $p_t = p + B_z^2/2\mu_0$ over
the vector potential $A$, which essentially requires that the same
field line be crossed twice by an observing spacecraft. Once the
invariant $z$ axis is acquired, the right-hand side of equation~6
can be derived from the differentiation of the best fit of $p_t$
versus $A$. This best fit is assumed to hold over the entire
flux-rope cross section. Away from the observation baseline, the
vector potential $A$ is calculated based on its second order Taylor
expansion with respect to $y$. Since the integration is
intrinsically a Cauchy problem, numerical singularities are
generated after a certain number of steps. As a result, the
transverse size is generally limited to half of the width along the
observation line in the integration domain. Detailed procedures can
be found in Hau and Sonnerup [1999] and Hu and Sonnerup [2002]. Here
we only use this approach to determine the axis orientation of MCs
and make a comparison with MVA.

\section{Lower Limits of the Transverse Extent}

\begin{figure*}
\noindent\includegraphics[width=12pc]{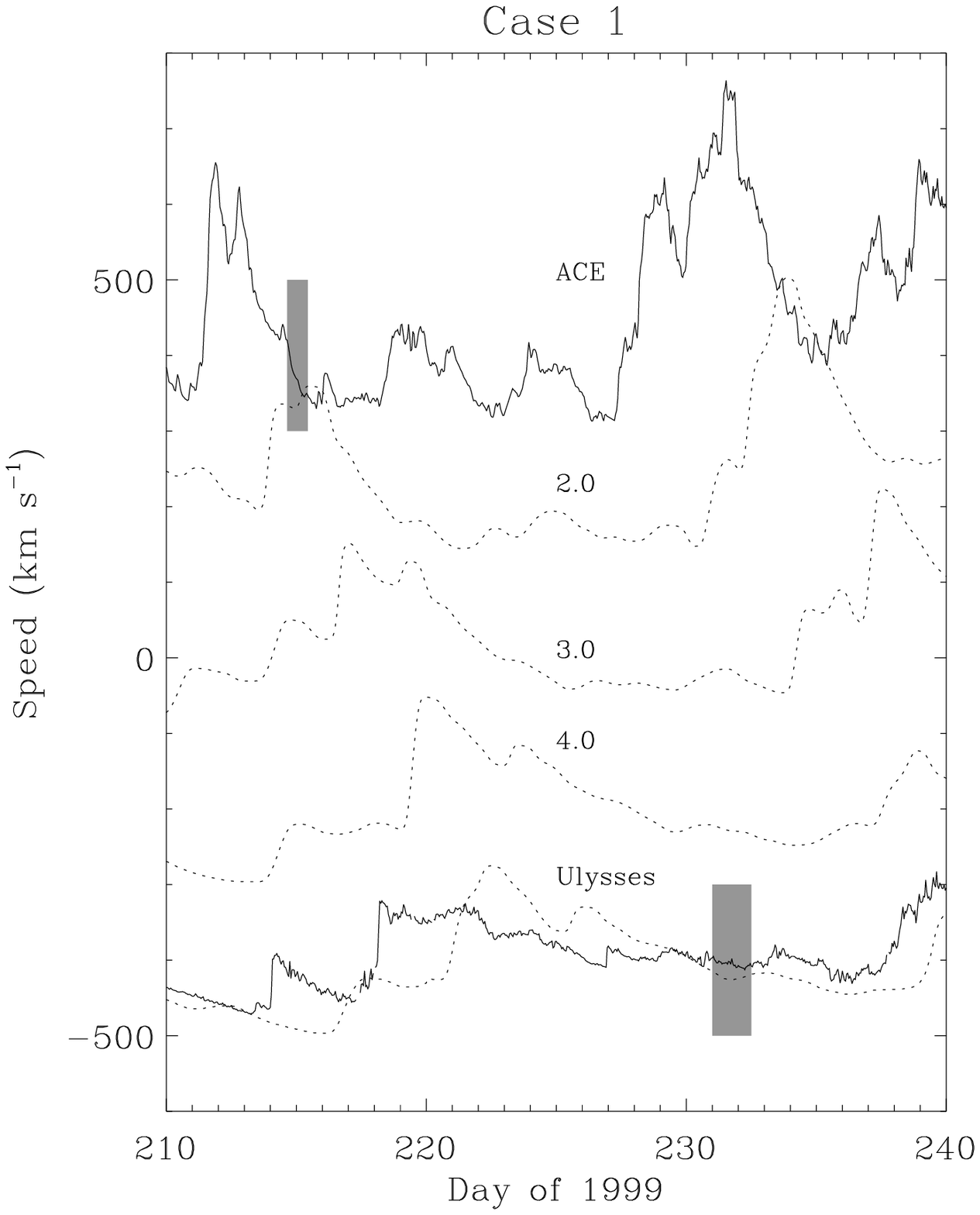}
\includegraphics[width=12pc]{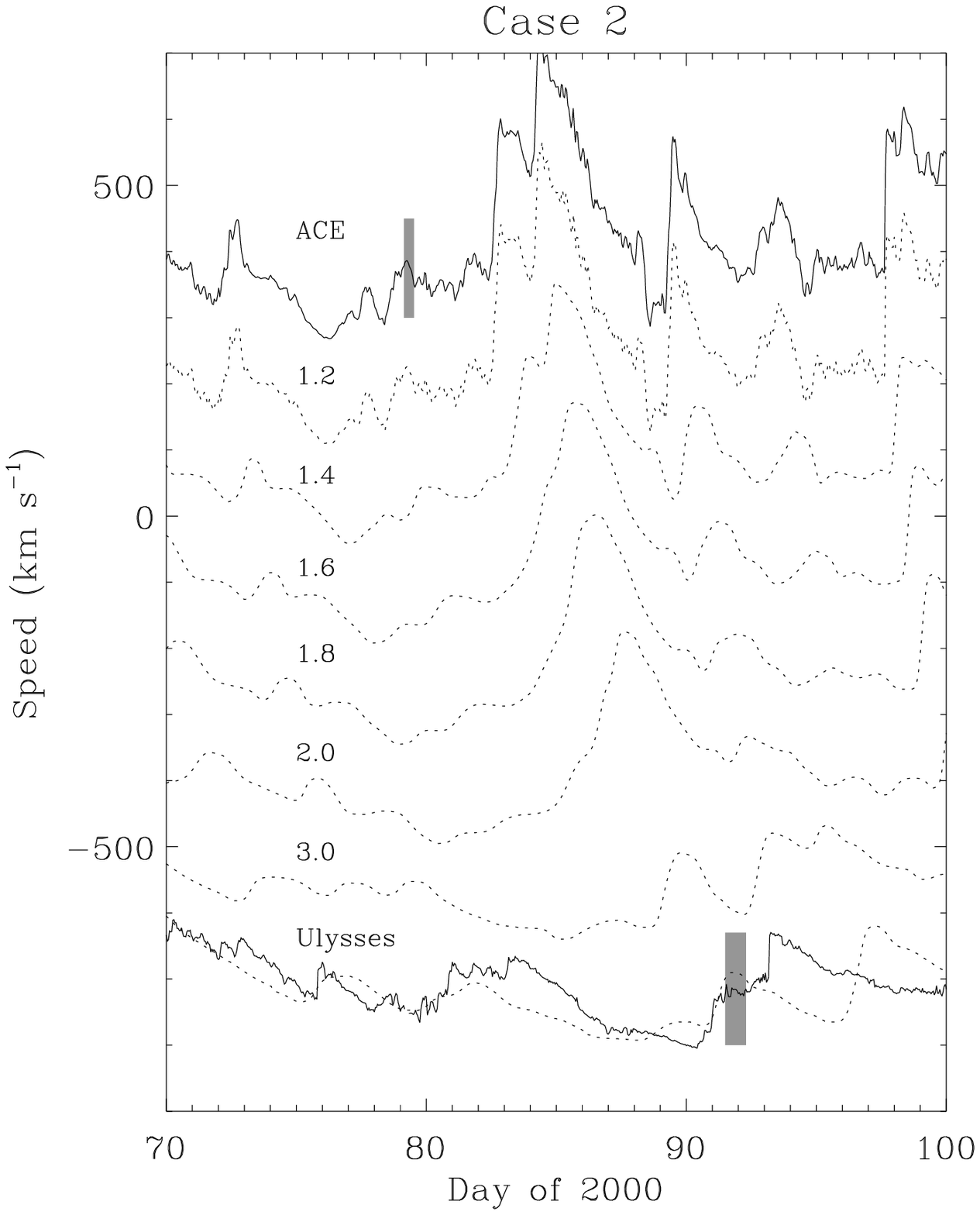}
\includegraphics[width=12pc]{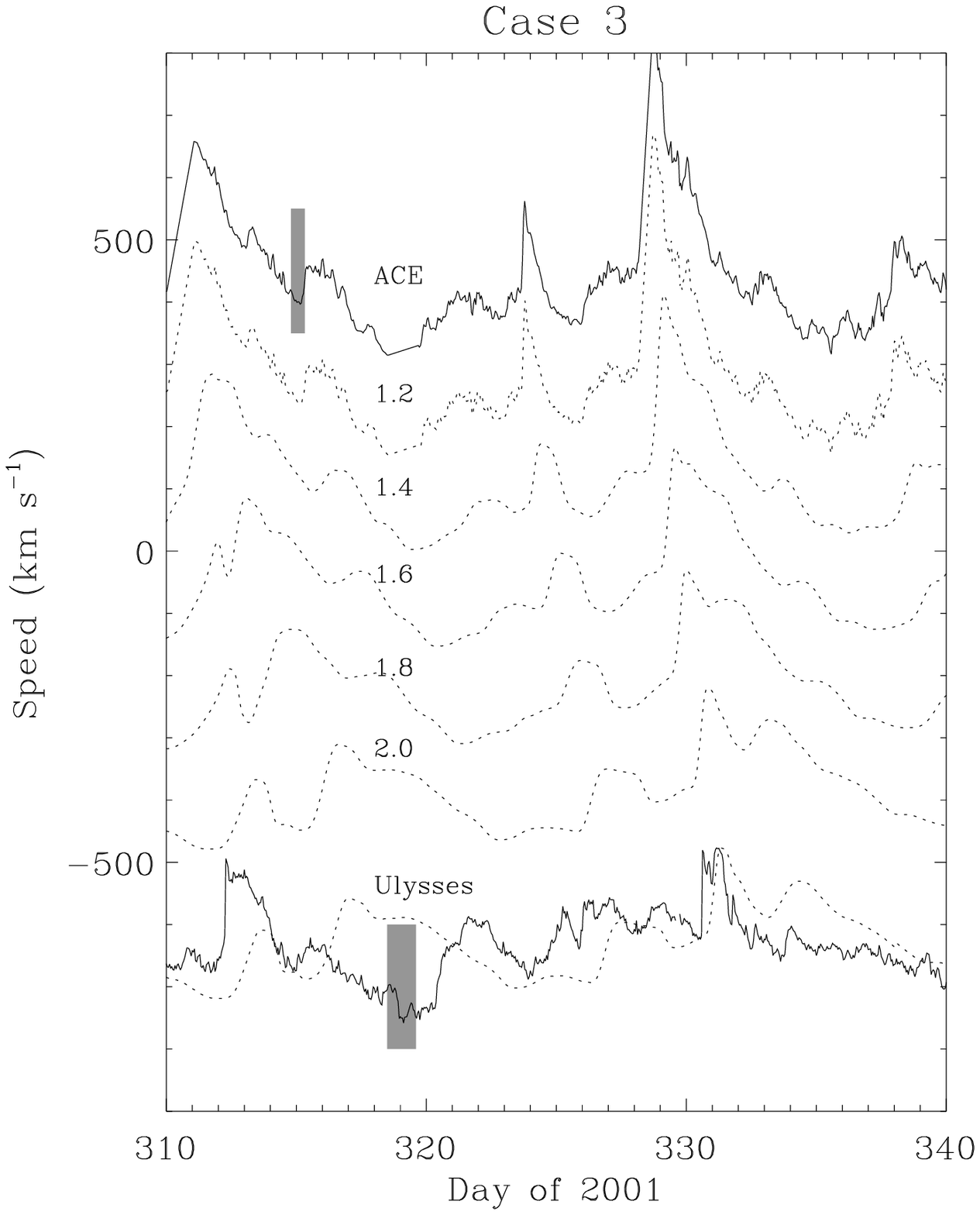}
\caption{Evolution of solar wind speed from ACE to Ulysses for the
three cases in Table~1 via the 1-D MHD model. The upper and lower
solid lines show the solar wind speeds observed at ACE and Ulysses,
while the dotted lines indicate the speed profiles predicted by the
model at distances (in AU) marked by the numbers. Shaded regions
represent the period where the MC was observed at ACE and Ulysses.
Each speed curve is decreased by 200 km s$^{-1}$ (left panel) and
160 km s$^{-1}$ (middle and right panels) with respect to the
previous one so that the individual line shapes can be easily
deciphered. For Case 3 (right panel), the profile at Ulysses is
shifted downward by 1360 km s$^{-1}$ from the observed speed, while
the model output at 2.3 AU is shifted by 1040 km s$^{-1}$ to line up
with the Ulysses data; the model output at Ulysses underestimates
the observed speed by 320 km s$^{-1}$.}
\end{figure*}

Application of the criteria and restrictions given in section~2
yields three MCs observed at both ACE and Ulysses with a latitudinal
separation larger than $30^{\circ}$. Table~1 lists the times,
locations, estimated axis orientations and chiralities for ACE and
Ulysses observations, respectively. The distance $R$, latitude
$\theta$ and longitude $\phi$ of the spacecraft are given in the
heliographic inertial frame. The MC denoted as PR in the table was
shown to be observed at both the spacecraft by Riley et al. [2003]
through a qualitative comparison between data and an MHD model
output; the two spacecraft were separated by $15^{\circ}$ in
latitude. Table~1 also gives the CMEs observed at the Sun (adopted
from http://cdaw.gsfc.nasa.gov/CME$_{-}$list) that best match the
occurrence time calculated from the MC's transit speed at 1 AU.
Applications of the MVA method to the normalized magnetic field
measurements and the GS technique to the plasma and magnetic field
observations within the MCs give the axis orientation, in terms of
the elevation ($\Theta$) and azimuthal ($\Phi$) angles. The axis
azimuthal angle is given in RTN coordinates (in which ${\bf R}$
points from the Sun to the spacecraft, ${\bf T}$ is parallel to the
solar equatorial plane and points to the planet motion direction,
and ${\bf N}$ completes the right-handed triad), which allows us to
see if the axis is perpendicular to the radial direction (i.e.,
close to 90$^{\circ}$ or 270$^{\circ}$). The magnetic field and
velocity vectors inside the MCs are rotated into the heliographic
inertial frame in order to have an axis elevation angle with respect
to the solar equatorial plane. The estimates of the axis orientation
from the MVA and GS methods roughly agree. The MCs generally lie
close to the solar equator and perpendicular to the radial
direction. The MC travel time from ACE to Ulysses is consistent with
the observed speeds. They also have the same chirality as listed in
Table~1.

Figure~2 shows the plasma and magnetic field measurements for Case 1
in Table~1 at ACE and Ulysses separated by 38$^{\circ}$ in latitude.
Its boundaries are mainly determined from the low proton temperature
combined with enhanced helium abundance. The helium enhancement has
been shown to be an effective tool to trace ICMEs from 1 AU to
Ulysses and Voyager 2 [e.g., Paularena et al., 2001; Richardson et
al., 2002]. The MVA of the magnetic field measurements inside the MC
at ACE and Ulysses gives an eigenvalue ratio
$\lambda_x:\lambda_y:\lambda_z$ of $6:1:0.8$ and $3.2:1:0.2$,
respectively. The MC has a relatively large axis elevation angle
($>40^{\circ}$) at ACE, probably due to the small separation between
$\lambda_y$ and $\lambda_z$; the axis orientation is close to the
solar equator at Ulysses. The bottom panels of Figure~2 display the
normalized magnetic field inside the MC projected onto the maximum
variance plane. The majority of the data show a coherent rotation of
about $180^{\circ}$; both the rotations indicate a left-handed
chirality of the magnetic configuration. Interestingly, the GS
reconstruction shows that this case contains two nested flux ropes
in both ACE and Ulysses measurements. Marked as a dashed line in
Figure~2, the separation of the two flux ropes in both ACE and
Ulysses observations appears as a discontinuity in the proton
density (second panels) and magnetic field components (sixth
panels). A shock wave, indicated by simultaneous sharp increases in
the proton density, speed, temperature and magnetic field strength,
follows the MC at ACE and Ulysses (not shown in the Ulysses data).

Figure~3 displays the data for Case 2 at ACE and Ulysses separated
by 43$^{\circ}$ in latitude. ACE and Ulysses are roughly aligned in
longitude for this case. The boundaries are determined from the
rotation of the magnetic field together with the low proton
temperature. This event is also associated with a small bump in the
helium/proton density ratio. The eigenvalue ratio
$\lambda_x:\lambda_y:\lambda_z$ determined from the MVA is
$2.5:1:0.2$ and $3.2:1:0.02$ for ACE and Ulysses observations,
respectively. It drives a forward shock at Ulysses which may also be
seen at ACE, but data gaps at ACE make it difficult to locate the
shock. Again, the normalized magnetic field data inside the MC at
ACE and Ulysses show a right-handed rotation in the maximum variance
plane.

Figure~4 shows the data for Case 3 which has the largest separation
(72$^{\circ}$) in latitude between ACE and Ulysses. Depressed proton
temperature and magnetic field rotation combined with enhanced
helium abundance are used to determine the boundaries. The velocity
and magnetic field fluctuations are strongly anti-correlated outside
the MC both at ACE and Ulysses, which indicates the presence of
Alfv\'{e}n waves; this wave activity also occurs within the MC but
at a reduced fluctuation level. The eigenvalue ratio
$\lambda_x:\lambda_y:\lambda_z$ from the MVA is $6.8:1:0.3$ for ACE
measurements and $3.4:1:0.1$ for Ulysses observations. Smooth
rotations of about 180$^{\circ}$ in the magnetic field (bottom
panels) indicate a left-handed chirality.

We propagate the ACE data to Ulysses using the 1-D MHD model
described in section~2. Figure~5 shows the velocity profiles
observed at ACE and Ulysses (solid lines), and the model profiles at
certain distances (dotted lines) for the three cases. In all of
these cases, larger streams at 1 AU persist to Ulysses as clearly
shown by the traces and model-data comparison at Ulysses, while
smaller ones smooth out. For Cases~1 and 2 (left and middle panels),
the model outputs at Ulysses agree qualitatively well with the
observed speed profiles. Note that the model-predicted and observed
speeds were not shifted to produce this agreement. For Case~3 (right
panel), the model predicts a slower solar wind than observed at
Ulysses, which is reasonable given the large latitudinal separation
(72$^{\circ}$) between ACE and Ulysses and possible differences in
the ambient solar wind. Based on the good stream alignment and data
similarities shown in Figures~2, 3 and 4, we conclude that the two
spacecraft see the same events.

The MCs listed in Table~1 offer observational evidence for the large
transverse size of ICMEs. In order to quantify the transverse size,
we examine the latitudinal separation between ACE and Ulysses for
these MCs. The latitudinal separation ($\Delta\theta$) serves as a
measure of the transverse size, $S_t$, expressed as
\begin{equation}
S_t = R\Delta\theta,
\end{equation}
where $R$ is the heliocentric distance of the MCs at Ulysses. This
equation explicitly assumes that the latitudinal extent of the MCs
is constant during their propagation through the solar wind. The
results for the present MCs are plotted in Figure~6. The transverse
size given by the above equation is much larger than the MCs' radial
width obtained from their average speed multiplied by the time
duration. As can seen from Figure~6, the largest aspect ratio is
$15.6:1$. The PR event has a ratio of $2.6:1$ since $\Delta\theta$
is only 15.3$^{\circ}$ for this case. Figure~6 reveals that the MCs
have a cross section greatly elongated in the latitudinal direction.

The large transverse size can also be inferred from the shock
standoff distance $d$ ahead of fast MCs written as [Russell and
Mulligan, 2002] $${d\over L} = 0.41 {(\gamma-1)M^2 + 2 \over
(\gamma+1)(M^2-1)},$$ where $\gamma={5\over 3}$, $M$ is the Mach
number of the preceding shock, and $L$ is the characteristic scale
of MCs, presumably a measure of $S_t$. From the superposed epoch
data of 18 near-Earth MCs with preceding shocks in Liu et al.
[2006b, Figure~8], we have $M=3.4$ and $d=0.17$ AU on average.
Substitution of these values into the above equation gives $L=1.2$
AU, which corresponds to a latitudinal extent of $\sim$ 69$^{\circ}$
obtained from equation~7. Consistent with our direct evidence, the
transverse size of MCs (or ICMEs in general) could be very large.

\section{Curvature of Magnetic Clouds}
A direct consequence of the large transverse size is that MCs
encounter different solar wind flows in the meridianal plane. MCs
can thus be highly distorted depending on the ambient solar wind
conditions. The simplified scenario described in section~1 indicates
that MCs should be ideally concave outward at solar minimum and
convex outward during solar maximum. This curvature effect results
in an inverse correlation between $\delta$ and $\theta$ at solar
minimum and a positive correlation near solar maximum as shown by
equation~1. Note, however, that this is a greatly simplified
picture. In reality, the shape of MCs will be determined by the
speed at which they travel with respect to the background solar
wind, ambient magnetic fields, the presence of other ICMEs or
obstacles nearby, and other features that are beyond the scope of
this paper.

As discussed in section~2, the distortion effect by solar wind flows
would be most prominent if MCs have axes close to the solar equator
and perpendicular to the radial direction. In order to have enough
events for our curvature analysis, we include all the MCs whose axes
lie within 30$^{\circ}$ of the solar equatorial plane and more than
30$^{\circ}$ away from the radial direction.

\subsection{MCs in a Structured Solar Wind}
Close to solar minimum, the solar wind is well ordered with fast
wind originating from polar coronal holes and slow wind near the
solar equatorial plane. Wind observations from 1995 - 1997 are used
to assess the distortion effect of the latitudinal flow gradient on
MCs. Figure~7 shows the normal elevation angles for the 14 events
observed at Wind. The error bars are obtained from equation~5. An
inverse correlation is observed between $\delta$ and $\theta$,
except for three events. The MCs are expected to be concave outward
during solar minimum, which is largely observed, made evident by the
inverse relationship plotted as a solid line. The three events where
$\delta$ and $\theta$ have the same sign indicate a convex outward
curvature, contrary to the solar minimum prediction. A closer look
at the 1997 May 15 event reveals an increasing speed profile
indicative of a high-speed stream interacting with the sunward edge
of the MC. As a result, the MC is bent to be convex outward. Note
that the simplified picture described in section~1 assumes a minimum
solar wind speed at the solar equator. It is conceivable that
$\delta$ and $\theta$ may have the same sign if the minimum speed
shifts away from the zero latitude. Although the breakdown of the
assumption is a possible explanation for the other two exceptions,
this effect may be negligible since most of the events in Figure~7
have the inverse correlation. The best fit to the data that show the
inverse correlation, obtained with a least squares analysis of
equation~1, gives a radius of curvature of $-0.3$ AU. Compared with
the convex-outward representation (dashed line), the data clearly
show the trend pictured by the right panel of Figure~1.

\subsection{MCs in a Uniform-Speed Solar Wind}
At solar maximum, the solar wind speed tends to be more uniform over
heliographic latitude. We use Ulysses observations between 1999 and
2003 to quantify the curvature effect of solar wind spherical
expansion on MCs. Our prescription yields 13 events shown in
Figure~8. In a uniform solar wind the MCs would have a radius of
curvature equal to their distance from the Sun, which results in a
proportional relationship, i.e., $\delta=\theta$ from equation~1.
Only two events do not have a positive correlation. More
specifically, $\delta$ and $\theta$ have opposite signs for the two
events. They are the first and last ones in the time series of the
MCs; as indicated by the dates, they may not truly come from the
solar maximum environment. The data nearby the $\delta=\theta$ curve
(dashed line) manifest a convex-outward structure at solar maximum
as illustrated by the left panel of Figure~1.

\section{Summary and Discussion}
We have investigated the transverse size and curvature of the MC
cross section, based on ACE, Wind and Ulysses observations. The
results provide compelling evidence that MCs are highly stretched in
the latitudinal direction and curved in a fashion depending on the
background solar wind.

Three MCs, whose axes are close to the solar equator and roughly
perpendicular to the radial direction, are shown to pass ACE and
Ulysses (widely separated in latitude) successively. The MVA method
combined with the GS reconstruction technique is used to determine
the axis orientation and the observations at ACE and Ulysses are
linked using a 1-D MHD model. The latitudinal separation between ACE
and Ulysses gives a lower limit to the MCs' transverse size. Varying
from 40$^{\circ}$ to 70$^{\circ}$, it reveals that the transverse
size can be very large. The flattened cross section is a natural
result of a flux rope subtending a constant angle as it propagates
from the Sun through the heliosphere (see Figure~1).

The radius of curvature is obtained from a simple relationship
between the MC normal elevation angle and the latitude of an
observing spacecraft. The curvature of MCs in the solar wind with a
latitudinal speed gradient at solar minimum differs from that in the
uniform-speed solar wind near solar maximum. At solar minimum, MCs
are bent concave-outward by the structured solar wind with a radius
of curvature of about 0.3 AU; at solar maximum, they tend to be
convex outward with the radius of curvature proportional to their
heliographic distance. The distortion of MCs, resulting from the
interaction with the ambient solar wind, is mainly a kinematic
effect since magnetic forces are dominated by the flow momentum
[Riley and Crooker, 2004].

Improvement of our knowledge of the global structure of MCs (or
generic ICMEs) is of critical importance for heliosphere physics and
space weather prediction. Proper estimates of the magnetic flux and
helicity of ICMEs require knowledge of the structure to quantify
their connection with the coronal origin and to assess the
modulation of heliospheric flux by CMEs. The large transverse size
and curvature can alter the global configuration of the
interplanetary magnetic field as ICMEs sweep through the
heliosphere. Numerical simulations show that the ambient magnetic
field extending from the Sun to high latitudes may bend poleward to
warp around the flattened flux rope [e.g., Manchester et al., 2004],
as initially proposed by Gosling and McComas [1987] and McComas et
al. [1988]. The field line draping leads to favorable conditions for
the formation of plasma depletion layers and mirror mode
instabilities in the sheath region of fast ICMEs [Liu et al.,
2006b]. A magnetic field bent southward would also allow for strong
coupling between the solar wind and the magnetosphere via field line
merging [Dungey, 1961]. The curvature of ICMEs could modify the
shape of preceding shock fronts, affecting plasma flows and particle
acceleration at the shocks.

\begin{figure}
\noindent\includegraphics[width=20pc]{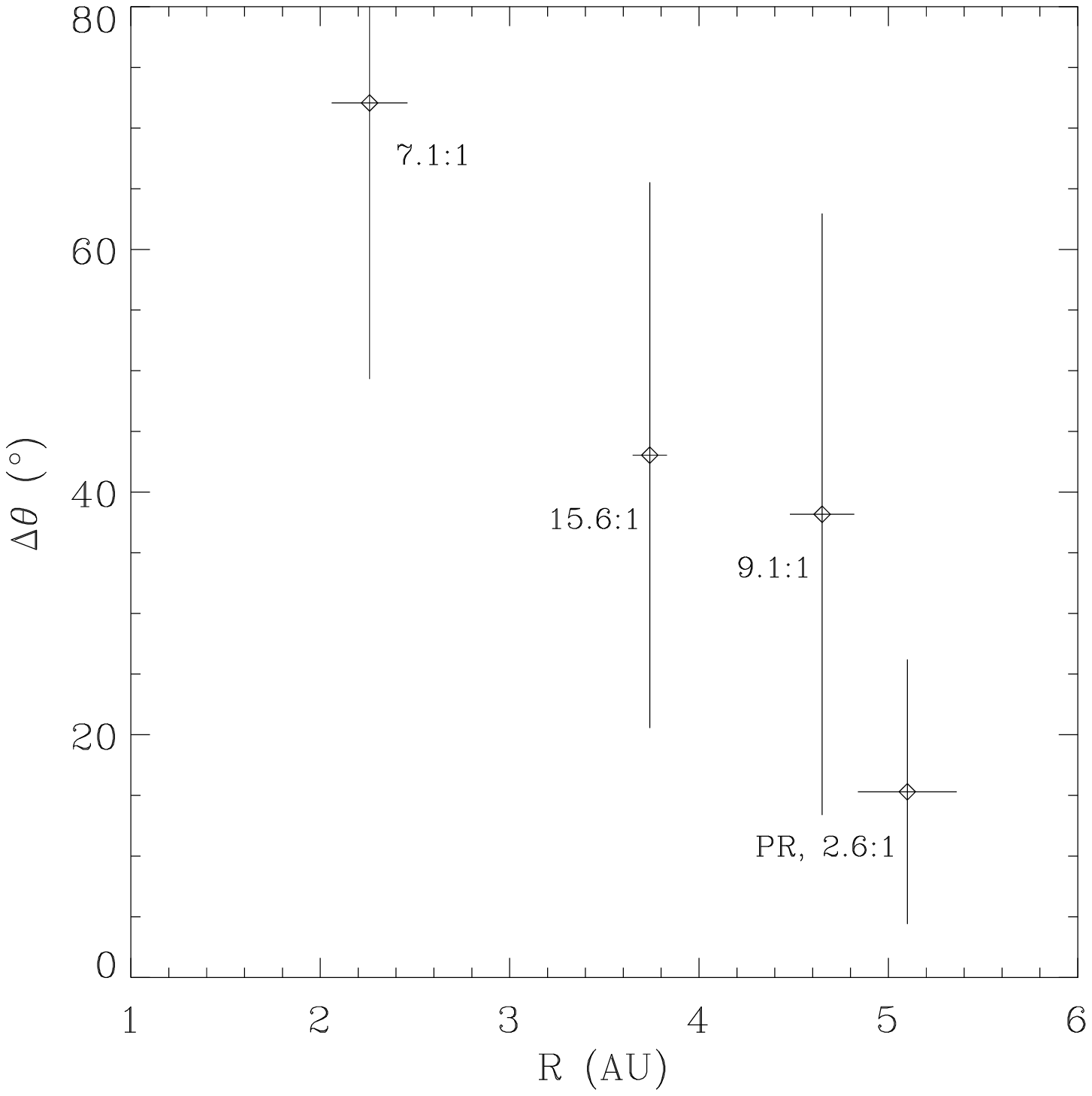} \caption{Latitudinal
separation between ACE and Ulysses for the MCs listed in Table~1 as
a function of Ulysses' heliocentric distance. The horizontal bars
show the radial width of the MCs, and the vertical bars indicate the
lower limit of the transverse size (converted to a length scale).
Text depicts the corresponding ratio of the two scales.}
\end{figure}

\begin{figure}
\noindent\includegraphics[width=20pc]{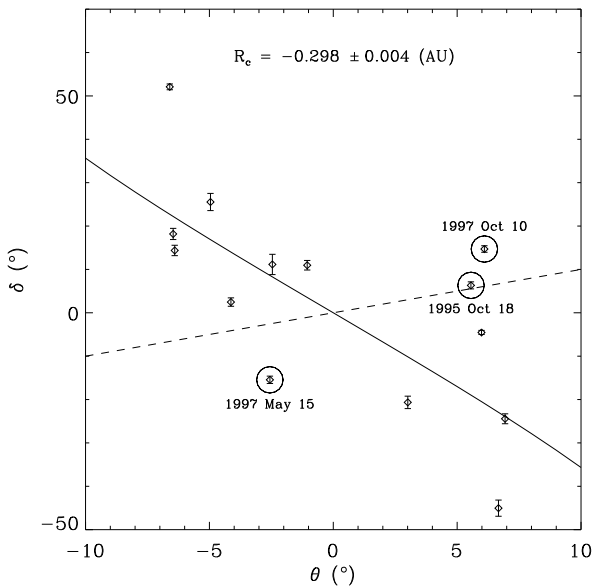} \caption{Elevation
angles of the MC normal from the solar equatorial plane as a
function of Wind's heliographic latitude. Circles with a nearby date
indicate events that do not have an inverse correlation. The solid
line represents the best fit of the data that obey the relationship
using equation~1. The radius of curvature resulting from the fit is
given by the text in the figure. The dashed line shows what would be
expected if the MCs were convex outward with a radius of curvature
of 1 AU.}
\end{figure}

\begin{figure}
\noindent\includegraphics[width=20pc]{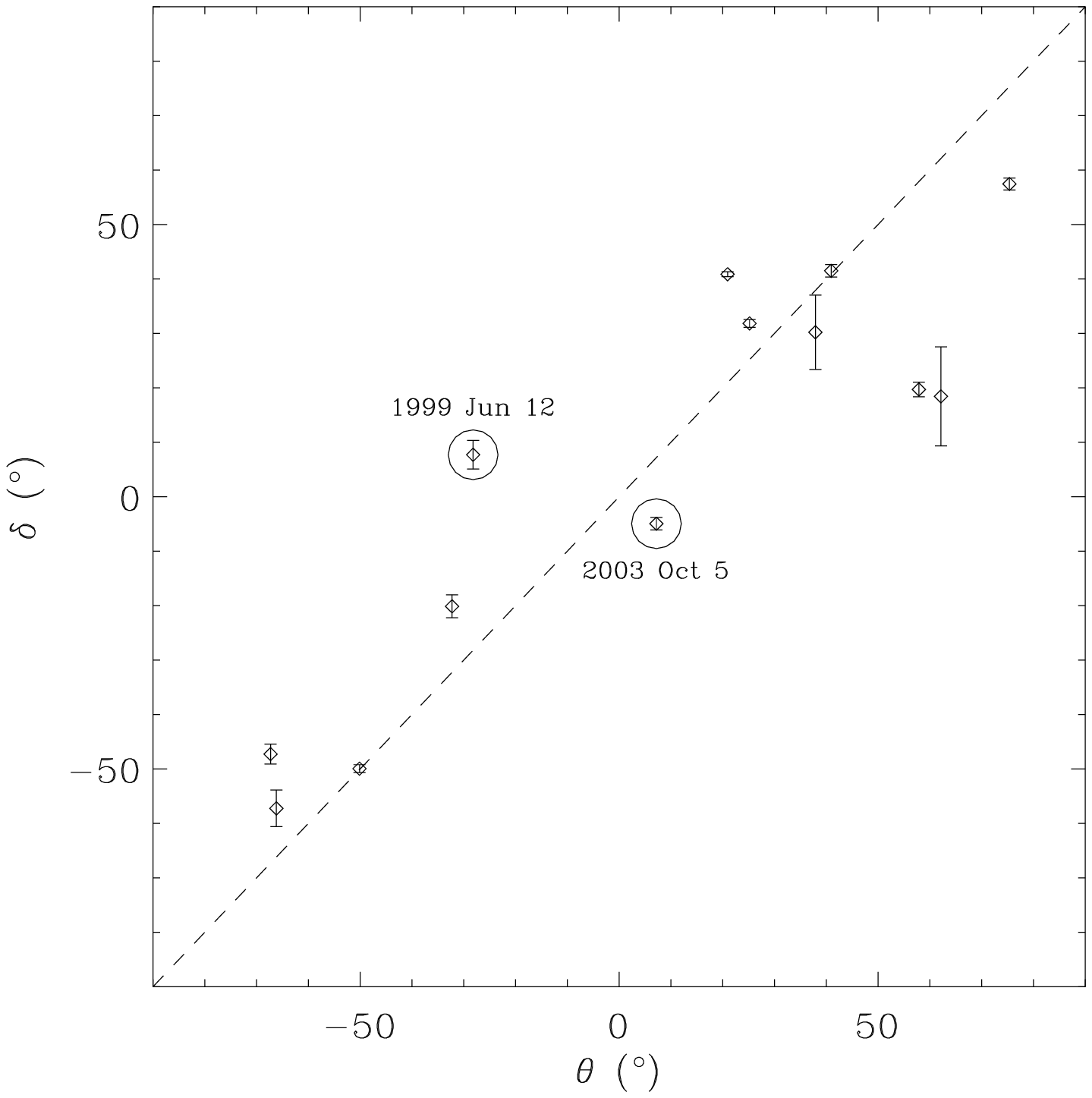} \caption{Elevation
angles of the MC normal from the solar equatorial plane as a
function of Ulysses' heliographic latitude. Circles with a nearby
date indicates events that do not have a positive correlation. The
dashed line represents $\delta=\theta$.}
\end{figure}

How general ICMEs are distorted remains unaddressed. Since their
magnetic field is not well organized, the difficulty resides in how
to best estimate their axis orientation. Future Stereo observations
will provide perspectives for their geometry and also quantitatively
test our results for MCs.

%
%

\begin{acknowledgments}
We acknowledge the use of ACE, Wind and Ulysses data from the NSSDC.
Y. L. thanks M. J. Owens of Boston University for helpful
discussion. The work at MIT was supported under NASA contract 959203
from JPL to MIT, NASA grants NAG5-11623 and NNG05GB44G, and by NSF
grants ATM-0203723 and ATM-0207775. This work was also supported in
part by the International Collaboration Research Team Program of the
Chinese Academy of Sciences.
\end{acknowledgments}

%
%

%
%
%

%
%

\end{article}


\begin{thebibliography}{}

\bibitem[]{}
Burlaga, L. F., E. Sittler, F. Mariani, and R. Schwenn (1981), Magnetic
loop behind an interplanetary shock: Voyager, Helios, and IMP 8
observations, \textit{J. Geophys. Res.}, \textit{86}, 6673.

\bibitem[]{}
Burlaga, L. F. (1988), Magnetic clouds and force-free fields with
constant alpha, \textit{J. Geophys. Res.}, \textit{93}, 7217.

\bibitem[]{}
Cargill, P. J., J. Schmidt, D. S. Spicer, and S. T. Zalesak (2000),
Magnetic structure of overexpanding coronal mass ejections:
Numerical models, \textit{J. Geophys. Res.}, \textit{105}, 7509.

\bibitem[]{}
Cid, C., M. A. Hidalgo, T. Nieves-Chinchilla, J. Sequeiros, and A.
F. Vi\~{n}as (2002), Plasma and magnetic field inside magnetic
clouds: A global study, \textit{Solar Phys.}, \textit{207}, 187.

\bibitem[]{}
Dasso, S., C. H. Mandrini, P. D\'{e}moulin, M. L. Luoni, and A. M.
Gulisano (2005), Large scale MHD properties of interplanetary
magnetic clouds, \textit{Adv. Space Res.}, \textit{35}, 711.

\bibitem[]{}
Dungey, J. W. (1961), Interplanetary magnetic field and the auroral
zones, \textit{Phys. Rev. Lett.}, \textit{6}, 47.

\bibitem[]{}
Gosling, J. T., and D. J. McComas (1987), Field line draping about
fast coronal mass ejecta: A source of strong out-of-ecliptic
interplanetary magnetic fields, \textit{Geophys. Res. Lett.},
\textit{14}, 355.

\bibitem[]{}
Gosling, J. T., D. J. McComas, J. L. Phillips, V. J. Pizzo, B. E.
Goldstein, R. J. Forsyth, and R. P. Lepping (1995), A CME-driven
solar wind disturbance observed at both low and high heliographic
latitudes, \textit{Geophys. Res. Lett.}, \textit{22}, 1753.

\bibitem[]{}
Groth, C. P. T., D. L. De Zeeuw, T. I. Gombosi, and K. G. Powell
(2000), Global three-dimensional MHD simulation of a space weather
event: CME formation, interplanetary propagation, and interaction
with the magnetosphere, \textit{J. Geophys. Res.}, \textit{105},
25,053.

\bibitem[]{}
Hammond, C. M., K. G. Crawford, J. T. Gosling, H. Kojima, J. L.
Phillips, H. Matsumoto, A. Balogh, L. A. Frank, S. Kokubun, and T.
Yamamoto (1995), Latitudinal structure of a coronal mass ejection
inferred from Ulysses and Geotail observations, \textit{Geophys.
Res. Lett.}, \textit{22}, 1169.

\bibitem[]{}
Hau, L.-N., and B. U. \"{O}. Sonnerup (1999), Two-dimensional
coherent structures in the magnetopause: Recovery of static
equilibria from single-spacecraft data, \textit{J. Geophys. Res.},
\textit{104}, 6899.

\bibitem[]{}
Hidalgo, M. A., C. Cid, A. F. Vi\~{n}as, and J. Sequeiros (2002a), A
non-force-free approach to the topology of magnetic clouds in the
solar wind, \textit{J. Geophys. Res.}, \textit{107}, 1002,
doi:10.1029/2001JA900100.

\bibitem[]{}
Hidalgo, M. A., T. Nieves-Chinchilla, and C. Cid (2002b), Elliptical
cross-section model for the magnetic topology of magnetic clouds,
\textit{Geophys. Res. Lett.}, \textit{29}, 1637,
doi:10.1029/2001GL013875.

\bibitem[]{}
Hu, Q., and B. U. \"{O}. Sonnerup (2002), Reconstruction of magnetic
clouds in the solar wind: Orientations and configurations,
\textit{J. Geophys. Res.}, \textit{107}, 1142,
doi:10.1029/2001JA000293.

\bibitem[]{}
Lepping, R. P., J. A. Jones, and L. F. Burlaga (1990), Magnetic
field structure of interplanetary magnetic clouds at 1 AU,
\textit{J. Geophys. Res.}, \textit{95}, 11,957.

\bibitem[]{}
Liu, Y., J. D. Richardson, and J. W. Belcher (2005), A statistical
study of the properties of interplanetary coronal mass ejections
from 0.3 to 5.4 AU, \textit{Plan. Space Sci.}, \textit{53}, 3,
doi:10.1016/j.pss.2004.09.023.

\bibitem[]{}
Liu, Y., J. D. Richardson, J. W. Belcher, J. C. Kasper, and H. A.
Elliott (2006a), Thermodynamic structure of collision-dominated
expanding plasma: Heating of interplanetary coronal mass ejections,
\textit{J. Geophys. Res.}, \textit{111}, A01102,
doi:10.1029/2005JA011329.

\bibitem[]{}
Liu, Y., J. D. Richardson, J. W. Belcher, J. C. Kasper, and R. M.
Skoug (2006b), Plasma depletion and mirror waves ahead of
interplanetary coronal mass ejections, arXiv:physics/0602164,
\textit{J. Geophys. Res.}, in press.

\bibitem[]{}
Manchester, W. B. IV, T. I. Gombosi, I. Roussev, A. Ridley, D. L. De
Zeeuw, I. V. Sokolov, K. G. Powell, and G. T\'{o}th (2004), Modeling
a space weather event from the Sun to the Earth: CME generation and
interplanetary propagation, \textit{J. Geophys. Res.}, \textit{109},
doi:10.1029/2003JA010150.

\bibitem[]{}
McComas, D. J., J. T. Gosling, D. Winterhalter, and E. J. Smith
(1988), Interplanetary magnetic field draping about fast coronal
mass ejecta in the outer heliosphere, \textit{J. Geophys. Res.},
\textit{93}, 2519.

\bibitem[]{}
McComas, D. J., et al. (1998), Ulysses' return to the slow solar
wind, \textit{Geophys. Res. Lett.}, \textit{25}, 1.

\bibitem[]{}
Mulligan, T., and C. T. Russell (2001), Multispacecraft modeling of
the flux rope structure of interplanetary coronal mass ejections:
Cylindrically symmetric versus nonsymmetric topologies, \textit{J.
Geophys. Res.}, \textit{106}, 10,581.

\bibitem[]{}
Odstrcil, D., J. A. Linker, R. Lionello, Z. Mikic, P. Riley, V. J.
Pizzo, and J. G. Luhmann (2002), Merging of coronal and heliospheric
numerical two-dimensional MHD models, \textit{J. Geophys. Res.},
\textit{107}, 1493, doi:10.1029/2002JA009334.

\bibitem[]{}
Odstrcil, D., P. Riley, and X. P. Zhao (2004), Numerical simulation
of the 12 May 1997 interplanetary CME event, \textit{J. Geophys.
Res.}, \textit{109}, doi:10.1029/2003JA010135.

\bibitem[]{}
Owens, M. J., V. G. Merkin, and P. Riley (2006), A kinematically
distorted flux rope model for magnetic clouds, \textit{J. Geophys.
Res.}, \textit{111}, doi:10.1029/2005JA011460.

\bibitem[]{}
Paularena, K. I., C. Wang, R. von Steiger, and B. Heber (2001), An
ICME observed by Voyager 2 at 58 AU and by Ulysses at 5 AU,
\textit{Geophys. Res. Lett.}, \textit{28}, 2755.

\bibitem[]{}
Richardson, J. D., K. I. Paularena, C. Wang, and L. F. Burlaga
(2002), The life of a CME and the development of a MIR: From the Sun
to 58 AU, \textit{J. Geophys. Res.}, \textit{107}, 1041,
doi:10.1029/2001JA000175.

\bibitem[]{}
Riley, P., J. A. Linker, Z. Miki, D. Odstrcil, T. H. Zurbuchen, D.
Lario, and R. P. Lepping (2003), Using an MHD simulation to
interpret the global context of a coronal mass ejection observed by
two spacecraft, \textit{J. Geophys. Res.}, \textit{108}, 1272,
doi:10.1029/2002JA009760.

\bibitem[]{}
Riley, P., J. A. Linker, R. Lionello, Z. Mikic, D. Odstrcil, M. A.
Hidalgo, C. Cid, Q. Hu, R. P. Lepping, B. J. Lynch, and A. Rees
(2004), Fitting flux ropes to a global MHD solution: A comparison of
techniques, \textit{J. Atmos. Solar-Terres. Phys.}, \textit{66},
1321.

\bibitem[]{}
Riley, P., and N. U. Crooker (2004), Kinematic treatment of coronal
mass ejection evolution in the solar wind, \textit{Astrophys. J.},
\textit{600}, 1035.

\bibitem[]{}
Russell, C. T. and T. Mulligan (2002), On the magnetosheath
thicknesses of interplanetary coronal mass ejections, \textit{Plan.
Space Sci.}, \textit{50}, 527.

\bibitem[]{}
Sonnerup, B. U. \"{O}, and Jr. L. J. Cahill (1967), Magnetopause
structure and attitude from Explorer 12 observations, \textit{J.
Geophys. Res.}, \textit{72}, 171.

\bibitem[]{}
Sonnerup, B. U. \"{O}, and M. Scheible (1998), Minimum and maximum
variance analysis, in \textit{Analysis Methods for Multi-Spacecraft
Data}, edited by G. Paschmann and P. W. Daly, pp. 185, Int. Space
Sci. Inst., Bern, Switzerland.

\bibitem[]{}
St. Cyr, O. C., et al. (2000), Properties of coronal mass ejections:
SOHO LASCO observations from January 1996 to June 1998, \textit{J.
Geophys. Res.}, \textit{105}, 18,169.

\bibitem[]{}
Sturrock, P. A. (Ed.) (1994), \textit{Plasma Physics: An
Introduction to the Theory of Astrophysical, Geophysical and
Laboratory Plasmas}, pp. 209, Cambridge Univ. Press, New York.

\bibitem[]{}
Vandas, M., and E. P. Romashets (2003), A force-free field with
constant alpha in an oblate cylinder: A generalization of the
Lundquist solution, \textit{Astron. Astrophys.}, \textit{398}, 801.

\bibitem[]{}
Wang, C., J. D. Richardson, and J. T. Gosling (2000), A numerical
study of the evolution of the solar wind from Ulysses to Voyager 2,
\textit{J. Geophys. Res.}, \textit{105}, 2337.

\bibitem[]{}
Wang, C., J. D. Richardson, and K. I. Paularena (2001), Predicted
Voyager observations of the Bastille Day 2000 coronal mass ejection,
\textit{J. Geophys. Res.}, \textit{106}, 13,007.

\bibitem[]{}
Webb, D. F., S. W. Kahler, P. S. McIntosh, and J. A. Klimchuck
(1997), Large-scale structures and multiple neutral lines associated
with coronal mass ejections, \textit{J. Geophys. Res.},
\textit{102}, 24,161.

\end{thebibliography}
\end{document}